\documentclass{article}

\usepackage[export]{adjustbox}
\usepackage{arxiv}
\usepackage{amsmath}
\usepackage[utf8]{inputenc} 
\usepackage[T1]{fontenc}    
\usepackage{hyperref}       
\usepackage{url}            
\usepackage{booktabs}       
\usepackage{amsfonts}       
\usepackage{nicefrac}       
\usepackage{microtype}      
\usepackage{lipsum}		
\usepackage{graphicx}
\usepackage{natbib}
\usepackage{doi}
\usepackage{xcolor}
\usepackage{siunitx}
\usepackage[justification=centering]{caption}

\title{Investigation of PINN Stability and Robustness for the Euler-Bernoulli Beam Problem}
\author{ \href{https://orcid.org/0000-0001-7802-7316}{\includegraphics[scale=0.06]{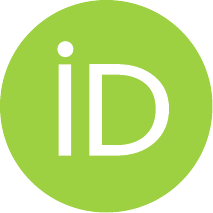}\hspace{1mm}Thonn~Homsnit} \\
	Mechanical Science Division,\\
	Saitama University\\
	\And
	\href{https://orcid.org/0000-0002-2262-6755}{\includegraphics[scale=0.06]{orcid.pdf}\hspace{1mm}Kensuke~Kageyama} \\
	Mechanical Science Division,\\
	Saitama University\\
    \And
	\href{https://orcid.org/0000-0003-0415-6451}{\includegraphics[scale=0.06]{orcid.pdf}\hspace{1mm}Tomohisa~Kojima}\thanks{Corresponding Author: \texttt{kojimat@mech.saitama-u.ac.jp}} \\
	Mechanical Science Division,\\
	Saitama University\\
}

\renewcommand{\headeright}{}
\renewcommand{\undertitle}{Preprint}
\renewcommand{\shorttitle}{Investigation of PINN Stability and Robustness for the Euler-Bernoulli Beam Problem}

\hypersetup{
pdftitle={A template for the arxiv style},
pdfsubject={q-bio.NC, q-bio.QM},
pdfauthor={Thonn~Homsnit, Kensuke~Kageyama, Tomohisa~Kojima}
}

\begin{document}
\maketitle

\begin{abstract}
	Physics-Informed Neural Networks (PINNs) encounter significant training difficulties when applied to doubly-clamped beam problems, and the underlying causes are not fully understood. This study investigates the PINN loss landscape to identify the failure mechanisms of two primary formulations: the high-order strong formulation and the energy-based formulation. The results demonstrate that the Strong Formulation suffers from landscape ill-conditioning driven by the boundary conditions (BCs), leading to convergence issues in the doubly-clamped case. Conversely, while the energy-based formulation requires only lower-order derivatives, its loss functional can become indefinite, causing optimization difficulties near saddle points. Based on strain field benchmarks against Finite Element Method (FEM), it is found that the strong formulation, combined with a BC handling method and the L-BFGS optimizer, yields the best performance across three classical boundary condition cases. These findings clarify distinct, formulation-dependent failure modes, offering a diagnostic foundation for developing robust physics-based surrogate models for complex beam systems.
\end{abstract}

\keywords{Physics-Informed Neural Network\and Euler-Bernoulli Beam \and Training Dynamics \and Structural Mechanics}

\section{Introduction}
Physics-Informed Neural Networks (PINNs) have been widely adopted for solving differential equations by injecting physics into the loss function \cite{raissi2019physics}. Since a PINN is a neural network with underlying physics, it offers a potential pathway for physics-based surrogate modeling. In solid mechanics applications, Haghighat et al. \cite{haghighat2021physics} studied data-assisted PINNs for plates and suggested that injecting physics improved robustness for the surrogate model. Additionally, Li et al. \cite{li2021physics} and Bastek et al. \cite{bastek2023physics}, who also worked with plate applications, investigated PINN behavior for field variable prediction using several differential equation formulations, such as strong-form, weak-form, and energy-based formulations. Bastek et al. \cite{bastek2023physics} found that, in their case, the strong form resulted in poor convergence due to the increased amount of competing penalty terms. Similarly, Li et al. \cite{li2021physics} mentioned that the weight of competing penalty terms needs further comprehensive investigation, as it is currently treated as a hyperparameter chosen by the user. Regarding the choice of differential equation formulation, Weinan and Yu \cite{weinan2018deep} noted that the Deep Ritz method, a variational formulation that eliminates penalty terms from boundary conditions (BCs), often encounters optimization landscapes characterized by saddle points and local minima.

In the field of beam applications, various PINN formulations have also been proposed to address specific structural problems. For example, Deep Reduced-Order Least-Square methods were utilized for Euler-Bernoulli and Timoshenko beams under various loading conditions \cite{luong2023deep}, and PINNs were applied to complex beam systems with simply supported conditions \cite{kapoor2023physics}. More recently, advanced formulations such as Variational PINNs (VPINN) have been developed to handle beams with elastic supports \cite{qian2025neural}. Additionally, systematic hyperparameter optimization has been investigated for the nonlinear analysis of nano-beams \cite{mirsadeghi2025physics}. Although these studies demonstrate the capability of PINNs in specific or modified scenarios, the general robustness of the standard PINN formulation across different boundary conditions, particularly the doubly clamped beam (CC), remains under-discussed. A detailed investigation into why standard PINNs become unstable specifically for CC beams has not yet been conducted.

The stability of the CC beam solution is of particular importance for the modeling of lattice structures and metamaterials. While current approaches for designing architected materials often rely on data-driven methods such as LSTMs \cite{lew2023designing}, a physics-based surrogate model is desirable for better generalization. Since lattice structures are mechanically composed of strut members that behave as CC beams \cite{ushijima2013prediction, tancogne2018stiffness}, the instability of PINNs in solving the individual CC beam problem presents a significant obstacle to constructing physics-based models for the entire structure.

In this paper, the instability of PINNs when applied to the Euler-Bernoulli CC beam problem is investigated. The focus is placed on the mechanics of the learning process and the loss landscape, analyzed using the Hessian in weight space, specifically examining the influence of loss formulation and optimizer choice on convergence. The limitations of the standard formulation in this specific boundary condition are identified to enable robust physics-based predictions.

\section{Materials and Method}
\subsection{Euler-Bernoulli Beam Problem}
\begin{figure}[ht]
    \centering
    \includegraphics[width=0.55\linewidth]{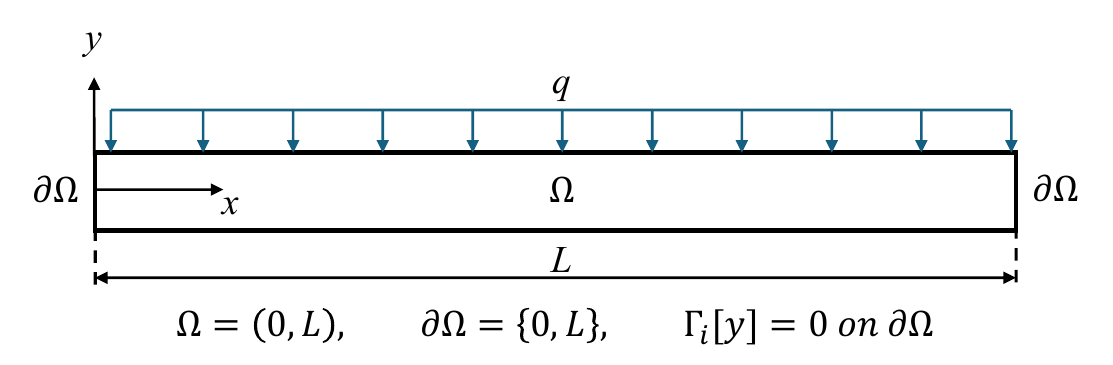}
    \caption{Euler-Bernoulli beam with distributed load and general boundary conditions.}
    \label{fig:1}
\end{figure}

In this study, we consider an Euler-Bernoulli beam subjected to a uniformly distributed load $q$ over its domain $\Omega$. The boundary conditions at both ends are illustrated in Figure \ref{fig:1}, where the boundary operator $\Gamma_i$ is applied to the deflection function to satisfy these conditions. We investigate three classical boundary conditions: cantilever (CV), simply-supported (SS), and doubly-clamped (CC).

The deflection $y(x)$ of this beam is governed by the following differential equation, assuming the beam is slender and undergoes small deflections:
\begin{equation}
    EI\frac{d^4y}{dx^4}=q 
\end{equation}
where $x \in [0,L]$ is the physical coordinate, $E$ is the elastic modulus, and $I$ is the moment of inertia.

Although this problem can be solved analytically, we employ PINNs to solve this differential equation to study their training dynamics for structural applications. PINNs typically operate in a normalized domain $\xi \in [0,1]$ to avoid saturation issues associated with continuous activation functions in the first layer, such as sigmoid and tanh. Therefore, the governing equation is nondimensionalized using the following variables.

The nondimensional variables and their corresponding characteristic values are defined as
\begin{equation}
    \xi=\frac{x}{L}, \qquad \hat y=\frac{y}{y_c}, \qquad \hat q=\frac{q}{q_c}
\end{equation}
where $L$, $y_c$, and $q_c$ are the characteristic length, deflection, and load, respectively. Substituting these variables into the physical governing equation and setting $y_c=\frac{q_cL^4}{EI}$ to balance the terms yields
\begin{equation}
    \frac{d^4\hat y}{d\xi^4}=\hat q
\end{equation}
This nondimensional strong formulation is used as the loss function for the PINN. However, the PINN loss function is not limited to the strong formulation; a variational or energy-based formulation, such as the potential energy functional, can also be employed. The nondimensional energy-based formulation is expressed as
\begin{equation}
    \Pi =\int_0^1\left({\frac{1}{2}\left(\frac{d^2\hat y}{d\xi^2}\right)^2-\hat q\hat y}\right)d\xi
\end{equation}
which implicitly satisfies the Neumann boundary conditions.

\subsection{PINN Formulation}
\begin{figure}[h]
    \centering
    \includegraphics[width=1.01\linewidth, trim={1.4cm  0 0 0},clip]{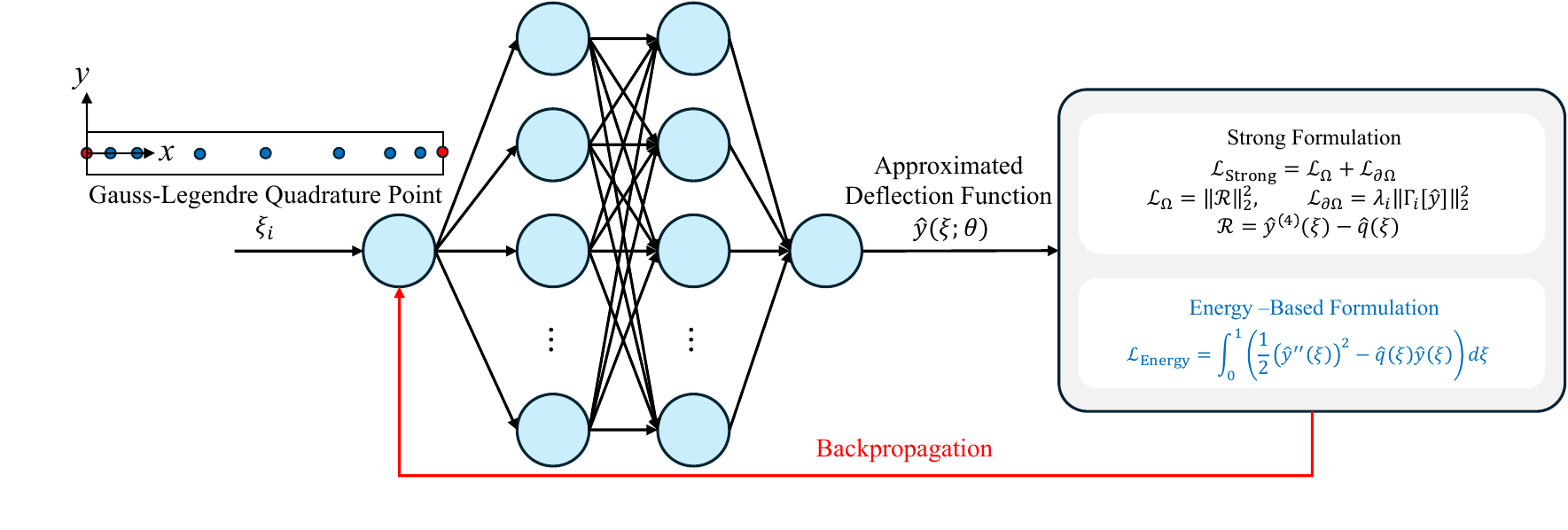}
    \caption{Schematic of the PINN architecture implementing Strong and Energy-based formulations.}
    \label{fig:2}
\end{figure}

The PINN architecture, illustrated in Figure \ref{fig:2}, is a deep neural network that approximates the deflection $\hat y$ as a function of the normalized coordinate $\xi$. The trainable parameters $\theta$ comprise the weights $\mathbf W$ and biases $\mathbf b$ across all layers. We employ a fully connected network with two hidden layers, 64 neurons per layer, and the hyperbolic tangent (tanh) activation function.

The network computes the deflection through a feed-forward process. The input coordinate $\mathbf y^{(0)}=\xi$ propagates through $N$ hidden layers, where the output $\mathbf y ^{(n)}$ of the $n$-th layer is given by:
\begin{equation}
    \mathbf y^{(n)}=\sigma\left( \mathbf W^{(n)} \mathbf y^{(n-1)} + \mathbf b^{(n)} \right), \qquad n=1,2,...,N
\end{equation}
Here, $\mathbf y ^{(n-1)}$ denotes the output of the preceding layer, $\sigma$ is the tanh activation function, and the final output $\mathbf y ^{(N)}$ represents the predicted deflection $\hat y$.

Training involves minimizing a loss function derived from the governing physics. Both the strong-form $L^2$-norm of the residual and the energy-based functional require integration over the domain, expressed as:
\begin{align}
    \mathcal L_{\mathcal R} &= \int_0^1(\mathcal R(\xi;\theta))^2d\xi \\
    \mathcal L_{\text{energy}} &= \int_0^1\left(\frac{1}{2} (\hat y''(\xi;\theta))^2-\hat q\hat y(\xi;\theta)\right)d\xi
\end{align}

We approximate these integrals using Gauss-Legendre (GL) quadrature. Since the tanh activation function is infinitely differentiable ($C^\infty$), GL quadrature provides high-accuracy numerical integration with a sparse set of sampling points, thereby reducing computational cost. The evaluated loss is used to iteratively update $\theta$ via backpropagation. We investigate training dynamics using both a first-order optimizer (Adam) and a quasi-Newton second-order optimizer (L-BFGS).

\section{Results and Discussions}
\subsection{Strong Formulation}

\begin{figure}[ht]
    \centering
    \includegraphics[width=1\linewidth]{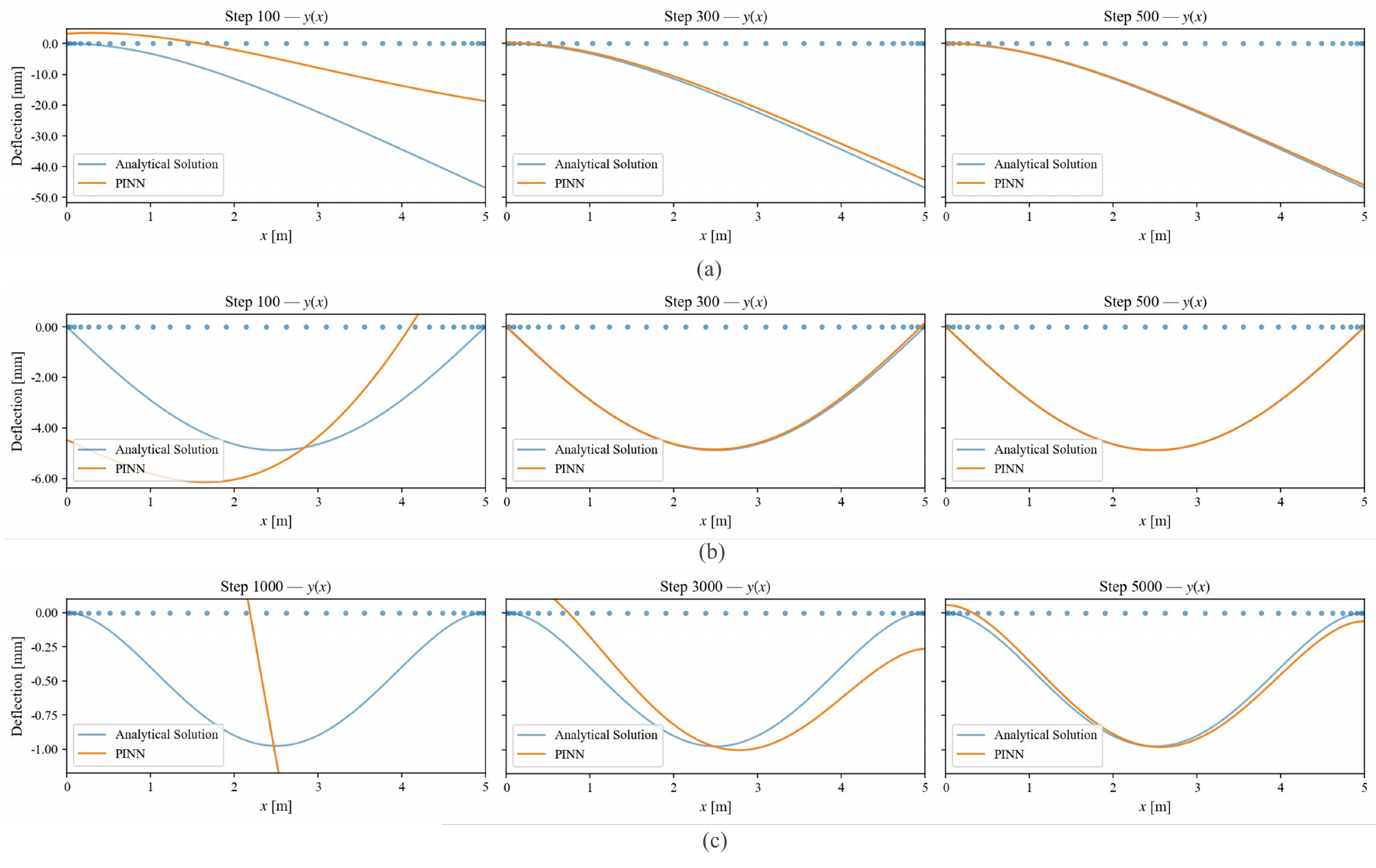}
    \caption{Snapshots of deflection predictions during training with Adam optimizer: (a) CV, (b) SS, and (c) CC.}
    \label{fig:3}
\end{figure}

We applied the Adam optimizer to all three cases for 5,000 iterations. As shown in Figures \ref{fig:3}(a) and \ref{fig:3}(b), the PINN rapidly converged to the analytical solution for both the CV and SS cases within 500 iterations. Meanwhile, Figure \ref{fig:3}(c) shows that the PINN with Adam failed to replicate the analytical solution for the CC beam, even after 5,000 iterations.

\begin{figure}[ht]
    \centering
    \includegraphics[width=1\linewidth]{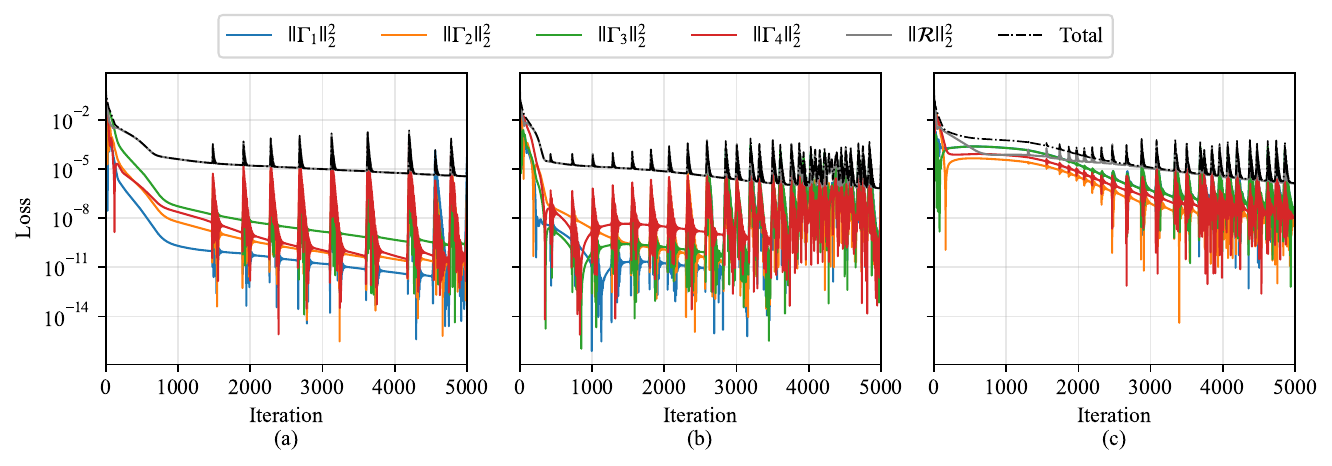}
    \caption{Evolution of individual loss components during training with Adam optimizer: (a) CV, (b) SS, and (c) CC.}
    \label{fig:4}
\end{figure}

The loss-iteration plots in Figure \ref{fig:4} provide insight into the failure in the CC case. For the successful CV and SS cases in Figures \ref{fig:4}a and \ref{fig:4}b, the BC penalty terms rapidly decrease and settle at a lower value than the residual loss. In contrast, for the CC case shown in Figure \ref{fig:4}c, the BC penalty terms dominate the residual loss for the first 2,000 iterations. This behavior demonstrates that the optimizer struggles to minimize the BC terms. Therefore, the loss landscape for the BC penalty in the CC case is difficult to minimize.

Moreover, the CV and SS losses exhibit instability. Loss spikes continuously occur even after convergence, which can cause the solution to drift away from the true solution. While an acceptable solution can be recovered by selecting the model state with the lowest loss, this instability demonstrates that using Adam is not robust for the CV and SS cases.

In addition, since the Euler-Bernoulli beam is a fourth-order ODE, its differential operator possesses a four-dimensional null space $\phi=[1,x,x^2,x^3]^T$ that is inherent to the approximated function from the PINN. Consequently, any network approximation $\hat y$ implicitly represents $\hat y_c$ from this null space:
\begin{equation}
    \hat y_c(\xi;\theta) =\phi(\xi)^T \mathbf c \qquad \text{where} \quad \mathbf c\in \mathbb R^4
\end{equation}

The loss in $L^2$-norm yields
\begin{equation}
    \mathcal{L}(c) = \sum_i^{4}||\Gamma_i[\hat y]||^2_2 = B_{ij}c_jB_{ik}c_k
\end{equation}
Therefore, the loss Hessian in the coefficient space is given by
\begin{equation}
    \mathbf{H}_{c}=\mathcal{L}_{,mn} = 2B_{im}B_{in} =2\mathbf{B}^T\mathbf{B}
\end{equation}

However, the PINN does not determine the implicit $\mathbf c $ directly. The implicit $\mathbf c$ is a function of the weights. Therefore, the Hessian of the loss in the weight space is
\begin{equation}
    \mathbf{H}_{\theta}=\mathcal{L}_{,\alpha\beta} = 2F_{m\alpha }B_{im}F_{k\beta }B_{ik}+2B_{im}B_{ik}c_k R_{m\alpha \beta}
\end{equation}
where $\frac{\partial c_i}{\partial \theta_\alpha}=F_{i\alpha}$ and $\frac{\partial^2 c_i}{\partial \theta_\alpha \partial \theta_\beta} =R_{i\alpha \beta}$. The second term can be omitted when the BC loss converges ($B_{ik}c_k \approx \textbf{0}$):
\begin{equation}
    \mathbf{H}_{\theta}\approx 2F_{m\alpha }B_{im}F_{k\beta }B_{ik} = 2(\mathbf{BF})^T(\mathbf{BF})=\mathbf{F}^T\mathbf{H}_{c}\mathbf{F}
\end{equation}

It can be seen that $\mathbf H_\theta$ inherits $\mathbf H_c$, but it also involves the mapping between the coefficient space and weight space via $\mathbf F$. Since $\mathbf B$ is row-wise linearly independent, $\mathbf H_c$ is symmetric positive definite (SPD). Therefore, $\mathbf B$ can be decomposed by singular value decomposition (SVD) as
\begin{equation}
    \mathbf B =\sum_i{\sigma_i [\mathbf u^{(i)}\otimes \mathbf v^{(i)}]}
\end{equation}
and
\begin{equation}
    \mathbf H_c =2\sum_i{\sigma^2_i[\mathbf v^{(i)}\otimes \mathbf v^{(i)}]}
\end{equation}
where
\begin{equation}
    \lambda(\mathbf H_c)=2\sigma(\mathbf B)^2
\end{equation}

Consequently, the condition number $\kappa(\mathbf H_c)$ can be calculated using $\frac{\lambda_{max}}{\lambda_{min}}$ since $\mathbf H_c$ is a normal matrix with 4 singular values $\text{diag}(2\tilde{\sigma})$. Applying the same decomposition strategy, $\mathbf H_\theta$ is given by:
\begin{align}
\mathbf{BF} &= \sum_{i} \tilde{\sigma}_{i}[\tilde{\mathbf{u}}^{(i)} \otimes \tilde{\mathbf{v}}^{(i)}]_{4 \times n_\theta} =
\begin{bmatrix}
\tilde{\mathbf{u}}^{(1)T} \\ \tilde{\mathbf{u}}^{(2)T} \\ \tilde{\mathbf{u}}^{(3)T} \\ \tilde{\mathbf{u}}^{(4)T}
\end{bmatrix}_{4 \times 4}^T
\begin{bmatrix}
\tilde{\sigma}_1 & 0 & 0 & 0 & \dots \\
0 & \tilde{\sigma}_2 & 0 & 0 & \dots \\
0 & 0 & \tilde{\sigma}_3 & 0 & \dots \\
0 & 0 & 0 & \tilde{\sigma}_4 & \dots
\end{bmatrix}_{4 \times n_\theta}
\begin{bmatrix}
\tilde{\mathbf{v}}^{(1)T} \\ \tilde{\mathbf{v}}^{(2)T} \\ \tilde{\mathbf{v}}^{(3)T} \\ \tilde{\mathbf{v}}^{(4)T} \\ \vdots
\end{bmatrix}_{n_\theta \times n_\theta}
\end{align}
Therefore,
\begin{align}
\mathbf{H}_\theta &= 2\sum_i{\tilde \sigma^2_i[\tilde{\mathbf v}^{(i)}\otimes \tilde{\mathbf v}^{(i)}]}_{n_\theta \times n_\theta} =
2\begin{bmatrix}
\tilde{\mathbf{v}}^{(1)T} \\ \tilde{\mathbf{v}}^{(2)T} \\ \tilde{\mathbf{v}}^{(3)T} \\ \tilde{\mathbf{v}}^{(4)T} \\ \vdots
\end{bmatrix}_{n_\theta \times n_\theta}^T
\begin{bmatrix}
\tilde{\sigma}_1^2 & 0 & 0 & 0 & \dots \\
0 & \tilde{\sigma}_2^2 & 0 & 0 & \dots \\
0 & 0 & \tilde{\sigma}_3^2 & 0 & \dots \\
0 & 0 & 0 & \tilde{\sigma}_4^2 & \dots \\
\vdots & \vdots & \vdots & \vdots & \ddots
\end{bmatrix}_{n_\theta \times n_\theta}
\begin{bmatrix}
\tilde{\mathbf{v}}^{(1)T} \\ \tilde{\mathbf{v}}^{(2)T} \\ \tilde{\mathbf{v}}^{(3)T} \\ \tilde{\mathbf{v}}^{(4)T} \\ \vdots
\end{bmatrix}_{n_\theta \times n_\theta}
\end{align}

\begin{figure}[h!]
    \centering
    \includegraphics[width=0.98\linewidth]{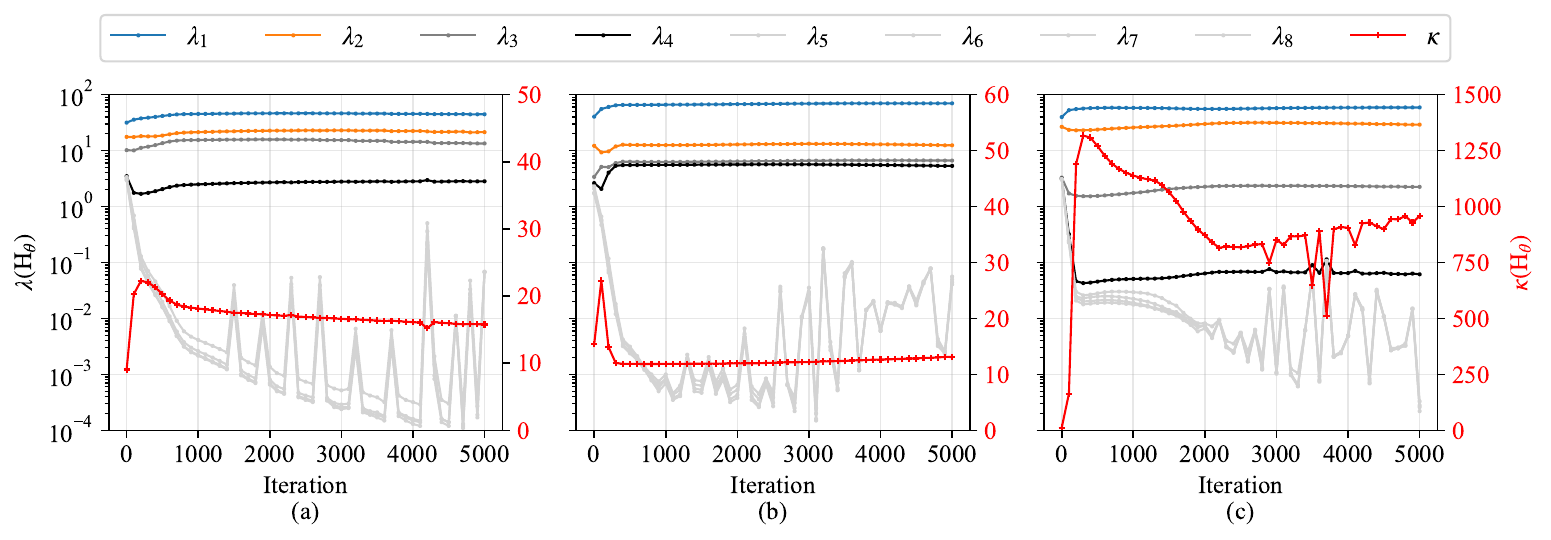}
    \caption{Evolution of the Hessian eigenvalues and condition number ($\kappa$) during training: (a) CV, (b) SS, and (c) CC.}
    \label{fig:5}
\end{figure}

It can be seen that $\mathbf H_\theta $ inherits only non-zero singular values from $\mathbf {BF}$. Therefore, there are only 4 active eigenvalues expected from $\mathbf H_\theta$. Figure \ref{fig:5} shows the evolution of these 4 active eigenvalues along with the non-active eigenvalues and the condition number. These results show that the loss Hessian of the BCs can gauge the severity of the ill-conditioning. We found that $\mathbf F$ occasionally acts as a preconditioner in the CV and SS cases, improving the condition number. However, in the CC case, where the condition number is approximately 10 times and 5 times higher than in the CV and SS cases, respectively, the condition number in the weight space worsens, as shown in Table \ref{tab:table1}, indicating that $F$ is not near-orthogonal. Therefore, the convergence problem in the CC case likely stems from the fact that the BCs make the loss landscape ill-conditioned. As a result, a first-order optimizer struggles to reach the minimum. Therefore, we employed the quasi-Newton second-order optimizer, L-BFGS, to address the ill-conditioned loss landscape of the CC case.

\begin{figure}[h!]
    \centering
    \includegraphics[width=0.95\linewidth]{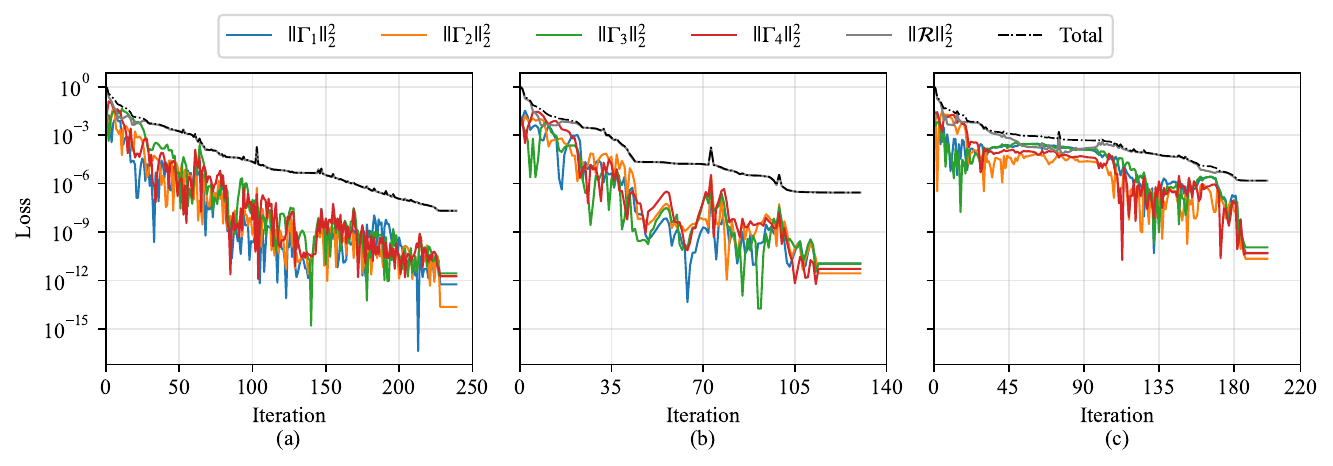}
    \caption{Evolution of individual loss components during training with L-BFGS optimizer: (a) CV, (b) SS, and (c) CC.}
    \label{fig:6}
\end{figure}

\begin{table}[h!]
    \caption{Comparison of condition numbers of Hessian in coefficient space ($\mathbf{H}_c$) and weight space ($\mathbf{H}_\theta$).}
    \label{tab:table1}
    \centering
    \setlength{\tabcolsep}{25pt}
    \begin{tabular}{lcc} 
    \toprule
    Case & $\kappa(\mathbf{H}_c)$ & $\kappa(\mathbf{H}_\theta)$ \\
    \midrule
    CV & 54.32 & 8.71 \\
    SS & 115.13 & 11.87 \\
    CC & 565.56 & 1158.35 \\
    \bottomrule
    \end{tabular}
\end{table}

After the L-BFGS optimizer was applied with 20 closure iterations, the optimizer took only around 200 backpropagations to successfully replicate the result in the CC case. The loss-iteration plots in Figure \ref{fig:6} show that L-BFGS not only improves the CC case but also resolves the loss instability problem in all three cases. Therefore, the choice of optimizer is crucial for PINNs that need to handle stiff BCs, which cause an ill-conditioned loss landscape. By utilizing the approximate inverse Hessian from L-BFGS, the ill-conditioned landscape becomes solvable.

\begin{figure}[h!]
    \centering
    \includegraphics[width=1\linewidth]{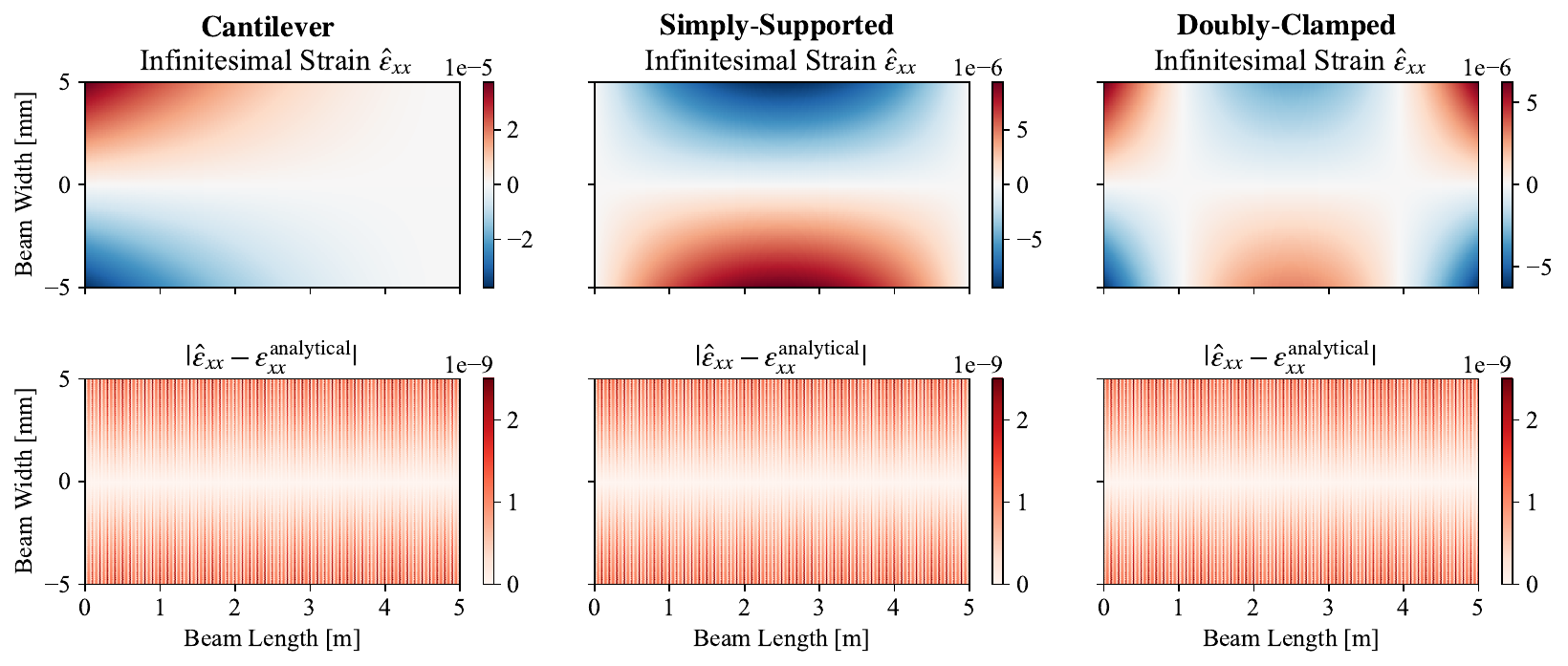}
    \caption{FEM benchmark results: Infinitesimal strain distribution (top) and absolute error (bottom).}
    \label{fig:7}
\end{figure}

Additionally, in solid mechanics problems, secondary field variables like strain and stress are critical, and errors from the deflection function can propagate to the strain field. Considering a linear elastic material, the error in stress is simply the Young's modulus times the strain error. Therefore, the strain field error represents the error in both the strain and stress fields. Figure \ref{fig:7} shows the strain field approximation from Finite Element Method (FEM) used for benchmarking the PINN performance. We used 50 elements of the 1D cubic element formulation to ensure slope continuity, since the infinitesimal strain from small deflection is calculated by $\varepsilon=-z\frac{d^2 y}{dx^2}$. The absolute error in Figure \ref{fig:7} is the difference in strain magnitude compared to the analytical solution. The stripes indicate discontinuities in the curvature approximation due to the choice of element shape function. However, considering applications that require beam elements, such as truss or lattice structures, the cubic element is sufficient for efficient computation and is used here as a benchmark.

\begin{figure}[h!]
    \centering
    \includegraphics[width=1\linewidth]{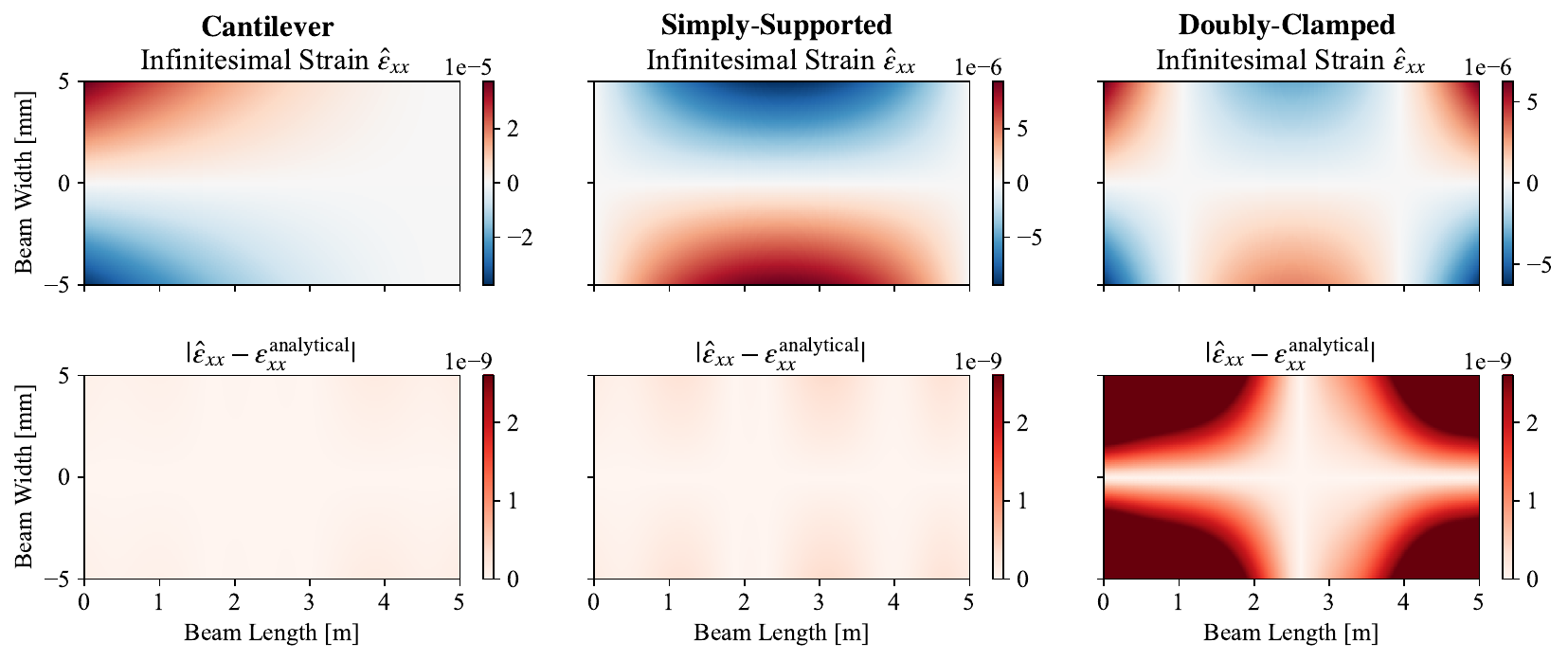}
    \caption{PINN predictions using the Strong Formulation optimized by L-BFGS:\\ Strain field distribution (top) and absolute error (bottom).}
    \label{fig:8}
\end{figure}
To qualitatively compare the strain field approximation, Figure \ref{fig:8} shows the strain field prediction from the PINN, which appears identical to the FEM approximation. However, if the same strain error cut-offs are used for comparison, it can be seen that the CV and SS cases not only achieve the benchmark but also produce a smoother strain approximation. Meanwhile, in the CC case, although it also yields a smoother strain field approximation, the absolute error exceeds the benchmark. Therefore, even though L-BFGS can mitigate the convergence problem in the CC case, the accuracy of the secondary field variable remains poor.

These results highlight a key weakness of combining the strong formulation with penalty terms. The optimizer is forced to balance the residual loss against BC loss. Moreover, even though a second-order optimizer can mitigate the convergence problem for ill-conditioned cases like the CC beam, the strain field error in all cases suggests that kinematic compatibility at the boundary is not fully guaranteed.

\subsection{BC-handling method}
To guarantee kinematic compatibility, we introduce an ansatz function that automatically satisfies the Dirichlet BCs, defined as $\hat y=f(\xi)N(\xi;\theta)$. Here, $f(\xi)$ represents the ansatz function, and $N(\xi;\theta)$ denotes the raw network output. The Hessian of the residual term in the weight space, $\mathbf H_{\mathcal R}$, is given by
\begin{equation}
    \mathbf H_{\mathcal R}=2\int_0^1\left(\left(\frac{\partial \mathcal R}{\partial \theta}\right)^2+\frac{\partial^2 \mathcal R}{\partial \theta^2} \right)d\xi\approx 2\int_0^1\left(\frac{\partial \mathcal R}{\partial \theta}\right)^2d\xi
\end{equation}
The residual Hessian is approximated via the Gauss-Newton method. Thus, the second-order term is omitted, ensuring the matrix is positive semi-definite (PSD). Considering the Hessians from both the residual and BC terms, the total Hessian is
\begin{align}
    \mathbf H _\text{total}&\approx2\int_0^1\left(f^{(4)}\frac{\partial N}{\partial \theta}+4f^{(3)}\frac{\partial N'}{\partial \theta}+6f''\frac{\partial N''}{\partial \theta}+ 4f'\frac{\partial N^{(3)}}{\partial \theta}+f\frac{\partial N^{(4)}}{\partial \theta}\right)^2d\xi +H_\theta^{(n)}
\end{align}
For the CV and SS cases, two Neumann BCs remain that are not incorporated into the ansatz function. Therefore, these two Neumann BC terms remain as penalty terms in the total Hessian:
\begin{equation}
    \mathbf H _\text{total(CV,SS)} \approx 2\int_0^1\left(6f''\frac{\partial N''}{\partial \theta}+4f'\frac{\partial N^{(3)}}{\partial \theta}+f\frac{\partial N^{(4)}}{\partial \theta}\right)^2d\xi +H_\theta^{(1)}+H_\theta^{(2)}
\end{equation}

Meanwhile, for the CC case, since all BCs are Dirichlet, they are already kinematically satisfied by the ansatz function. Consequently, no penalty terms remain. This results in a significantly better-conditioned loss landscape compared to the standard strong formulation with ill-conditioned penalty terms:
\begin{equation}
    \mathbf H _\text{total(CC)} \approx 2\int_0^1\left(f^{(4)}\frac{\partial N}{\partial \theta}+4f^{(3)}\frac{\partial N'}{\partial \theta}+6f''\frac{\partial N''}{\partial \theta}+4f'\frac{\partial N^{(3)}}{\partial \theta}+f\frac{\partial N^{(4)}}{\partial \theta}\right)^2d\xi
\end{equation}

\begin{figure}[h!]
    \centering
    \includegraphics[width=0.55\linewidth]{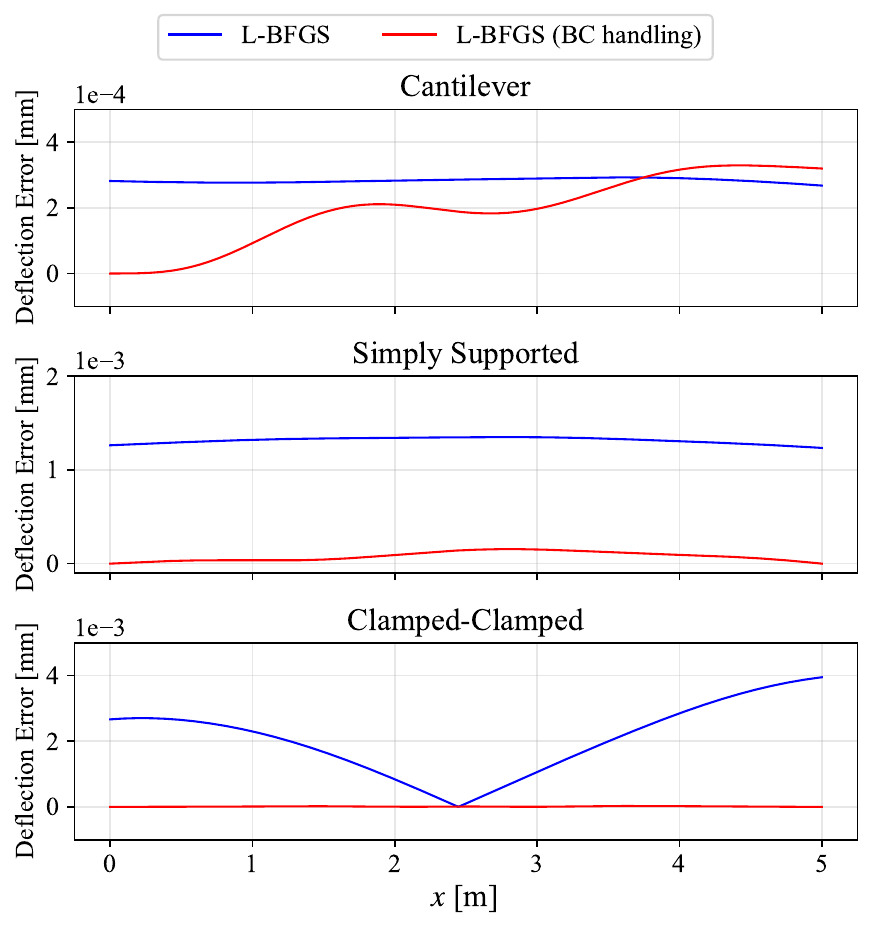}
    \caption{Deflection error profiles: Comparison between standard penalty and BC-handling formulations.}
    \label{fig:9}
\end{figure}

\begin{figure}[h!]
    \centering
    \includegraphics[width=0.65\linewidth]{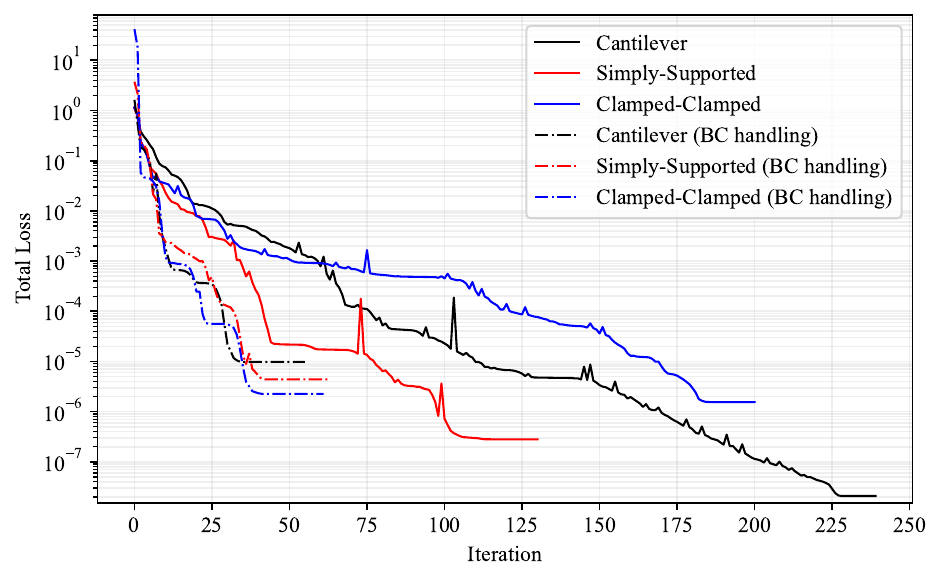}
    \caption{Comparison of total loss evolution between the standard formulation (solid lines)\\ and the BC-handling method (dashed lines).}
    \label{fig:10}
\end{figure}

Fig. \ref{fig:9} reveals that the standard penalty method fails to satisfy kinematic compatibility, despite achieving a lower total loss (see Fig. \ref{fig:10}). In contrast, the proposed BC-handling method enforces Dirichlet BCs via ansatz functions. Although this results in a slightly higher residual loss, it effectively eliminates kinematic incompatibilities, ensuring physically accurate deflections where boundary errors are strictly zero.

\begin{figure}[h!]
    \centering
    \includegraphics[width=1\linewidth]{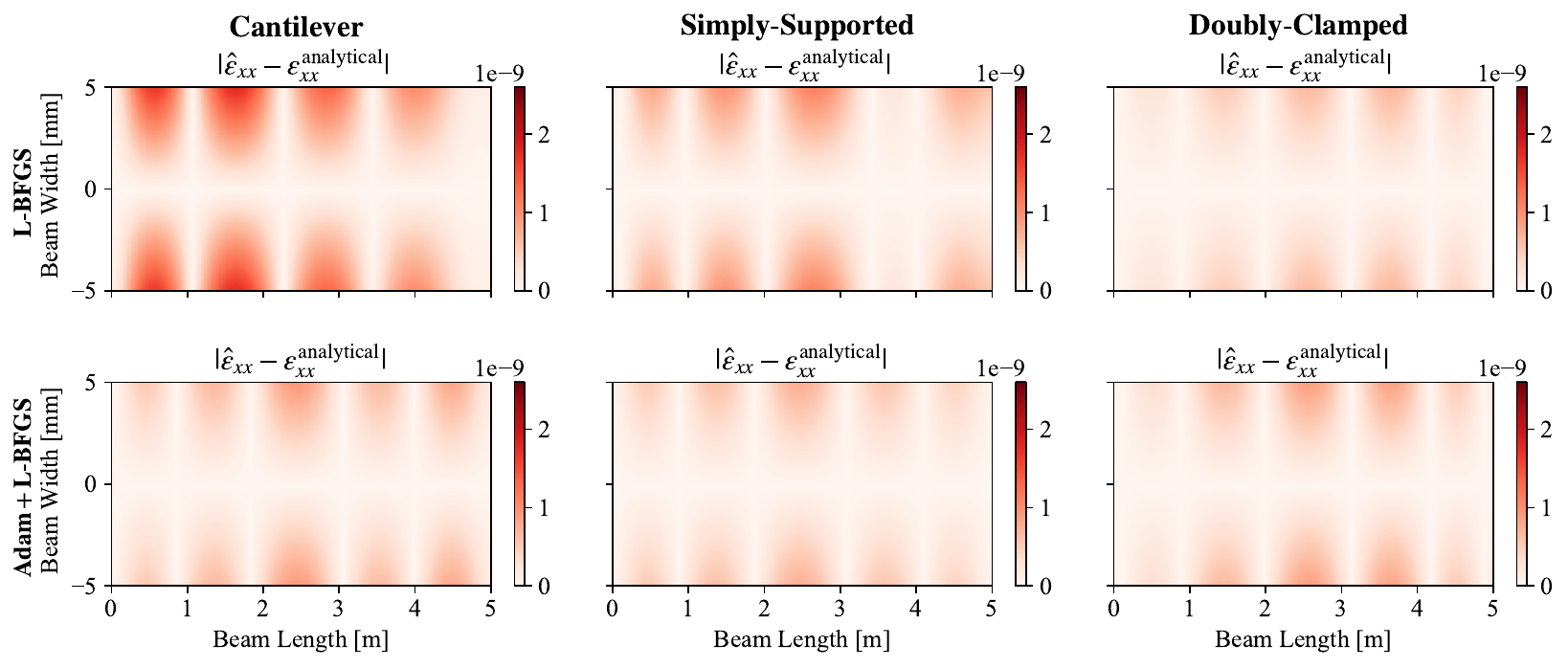}
    \caption{Strain field absolute errors using the BC-handling formulation:\\ L-BFGS optimizer (top) and Hybrid optimizer (bottom)}
    \label{fig:11}
\end{figure}

Moreover, since the CC case now exhibits a better-conditioned loss landscape, the strain field prediction is also improved, as shown in Fig. \ref{fig:11}. However, there is a trade-off for the CV and SS cases, as the standard penalty method already offers a well-conditioned loss landscape. Additionally, we employed a hybrid optimizer strategy, using an Adam warmup for the first 400 iterations followed by L-BFGS for another 100 closure iterations. This method yields a more consistent error magnitude across all three cases.

\subsection{Energy-Based Formulation}
The energy-based formulation represents another technique for PINNs, applying the minimum potential energy principle as the loss function, as the optimizer inherently seeks the loss minimum. In this study, we used the nondimensional form of the loss function:

\begin{equation}
    \mathcal{L}_{w} = C \int_{\Omega} \left(\frac{1}{2}(\hat y'')^2 - \hat q \hat y\right) d\xi \quad \text{where } C = \frac{q_c^2L^5}{EI}
\end{equation}

It can be seen that $C$ ensures consistency with physical energy units. However, to utilize the nondimensional form consistent with the strong formulation, only the integral term is considered, omitting $C$.

Considering the Hessian of the energy-based loss, Eq. \ref{eq:energy} reveals that the Hessian can potentially exhibit indefiniteness. This arises from the non-vanishing terms $\hat y''$ and $\hat q$, which involve the Hessians of the curvature and the deflection itself. Consequently, these terms cannot be neglected as in the Gauss-Newton approximation used for the strong formulation.

\begin{equation}
    \mathbf H_w =  \underbrace{ \int_{\Omega}  (\frac{\partial \hat y''}{\partial \theta})^2 d\xi}_\text{{PSD}} + \underbrace{\int_{\Omega} \left((\hat y'')\frac{\partial^2 \hat y''}{\partial \theta^2} -( \hat q) \frac{\partial^2 \hat y}{\partial \theta^2} \right) d\xi}_{\text{Potentially Indefinite}}
    \label{eq:energy}
\end{equation}

\begin{figure}[h!]
    \centering
    \includegraphics[width=0.65\linewidth, trim={0 0 9.3cm 0},clip]{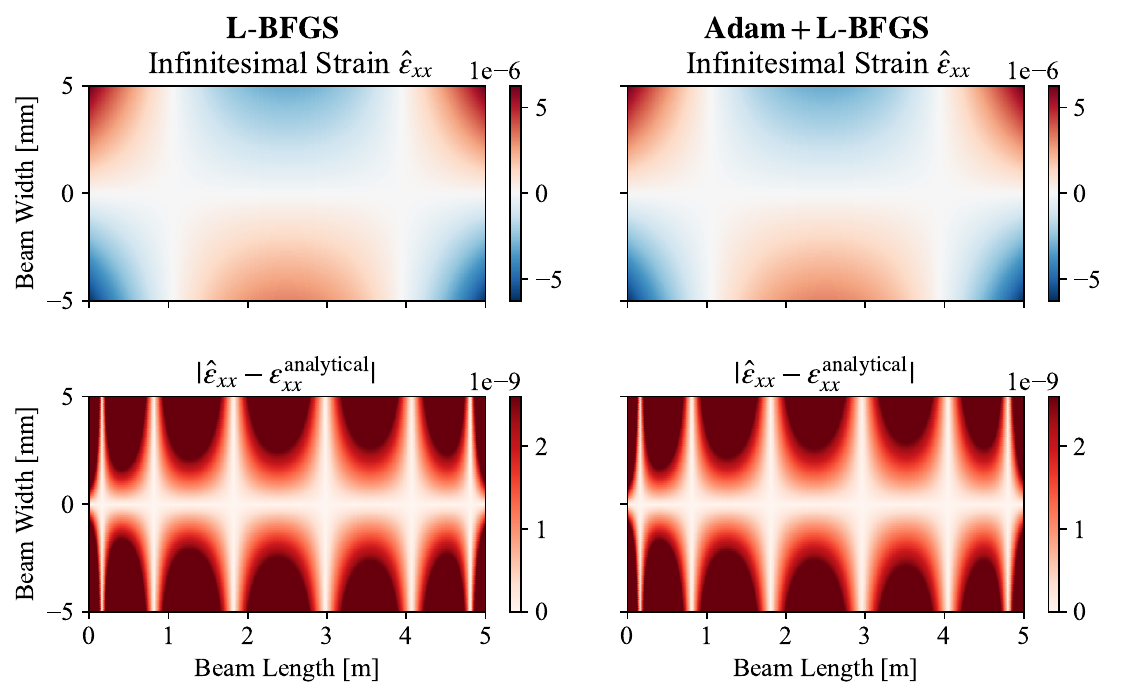}
    \caption{Strain field absolute errors using the Energy-based formulation: L-BFGS (left) and Hybrid optimizer (right).}
    \label{fig:12}
\end{figure}

Therefore, based on the sufficient condition for a saddle point, the potential indefiniteness of the Hessian implies the existence of stationary points where the optimizer may stagnate. This prevents the determination of the true minimizer. Fig. \ref{fig:12} compares the strain field prediction and error using L-BFGS and the hybrid optimizer. It is observed that the solutions are identical regardless of the optimizer choice. This consistency suggests that the optimization is trapped at the same stationary point, likely due to the indefinite nature of the Hessian creating a non-convex landscape dominated by saddle points or sub-optimal local minima.

To provide a quantitative comparison across all investigated methods, the maximum absolute strain errors relative to the analytical solution are summarized in Table \ref{tab:table2}. The results highlight that while the standard strong formulation and the energy-based approach suffer from significant errors in the stiff CC case, the strong-form BC-handling method with the hybrid optimizer consistently maintains high accuracy across all boundary conditions. Notably, it surpasses the FEM benchmark accuracy while offering the distinct advantage of a smooth, continuous field prediction inherent to the neural network function.

\begin{table}[h!]
	\caption{Comparison of maximum absolute strain errors relative to the \\analytical solution across all investigated methods.}
	\centering
    \begin{tabular}{llll} 
    \toprule
    Model & \multicolumn{1}{c}{CV} & \multicolumn{1}{c}{SS} & \multicolumn{1}{c}{CC} \\
    \midrule
    FEM & \textbf{$2.5 \times 10^{-9}$} & \textbf{$2.5 \times 10^{-9}$} & \textbf{$2.5 \times 10^{-9}$} \\
    Strong Form (L-BFGS)& \textcolor{olive}{$1.5 \times 10^{-10}$} & \textcolor{olive}{$3.6 \times 10^{-10}$} & \textcolor{red}{$8.3 \times 10^{-9}$} \\
    Strong Form (BC-embed, L-BFGS) & \textcolor{olive}{$1.8 \times 10^{-9}$} & \textcolor{olive}{$1.1 \times 10^{-9}$} & \textcolor{olive}{$6.8 \times 10^{-10}$} \\
    Strong Form (BC-embed, Hybrid) & \textcolor{olive}{$9.3 \times 10^{-10}$} & \textcolor{olive}{$7.6 \times 10^{-10}$} & \textcolor{olive}{$8.8 \times 10^{-10}$} \\
    Weak Form (BC-embed, L-BFGS)& \qquad - & \qquad - & \textcolor{red}{$1.9 \times 10^{-8}$} \\
    Weak Form (BC-embed, Hybrid)& \qquad - & \qquad - & \textcolor{red}{$1.9 \times 10^{-8}$} \\
    \bottomrule
    \end{tabular}
	\label{tab:table2}
\end{table}

\section{Conclusion}

In this study, we investigated the capability of PINNs to predict the deflection and strain fields of Euler-Bernoulli beams under three classical boundary conditions: CV, SS, and CC. The key findings are summarized as follows:

\begin{enumerate}
    \item The condition number of the loss landscape in the weight space inherits the same trend as the condition number of the boundary condition (BC) Hessian across all three cases.

    \item The strong formulation using the standard penalty method suffers from the stiffness of the BCs themselves. However, the choice of optimizer significantly affects predictive capability. By employing a second-order optimizer like L-BFGS, the PINN successfully replicates the analytical solution without instability or convergence issues. Nevertheless, it predicts the strain field with poor accuracy and fails to meet the FEM benchmark.

    \item With the BC-handling method, kinematic compatibility is strictly satisfied, and the strain field prediction for the CC case is significantly improved. Moreover, with the hybrid optimizer strategy, the maximum strain error is brought to a consistent magnitude across all cases, as shown in Table \ref{tab:table2}.

    \item The energy-based formulation exhibits potential indefiniteness in its Hessian. This can lead to saddle points in the loss landscape. Unlike the strong formulation, the second-order derivatives cannot be neglected via the Gauss-Newton approximation due to the existence of non-vanishing terms (curvature and load).
\end{enumerate}

These findings reveal that the strong formulation combined with the BC-handling technique and a hybrid optimizer strategy is the best practice for Euler-Bernoulli beam prediction. This is particularly important for modeling lattice structures, as they are mechanically composed of CC struts. Therefore, this work lays a foundation for developing physics-based surrogate models for lattice structures and complex beam systems, serving as a reliable alternative to statistical black-box tools.

\section*{Acknowledgments}
The first author acknowledges the financial support from the Japanese Government (MEXT) Scholarship. The authors also acknowledge the use of Google Gemini for assistance with English language editing and grammatical corrections.

\bibliographystyle{unsrt}
\bibliography{references}  
\end{document}